# Optimizing Native Ion Mobility Q-TOF in Helium and Nitrogen for Very Fragile Noncovalent Structures

Valérie Gabelica,[1*] Sandrine Livet[1‡] and Frédéric Rosu[2]

[1] Université de Bordeaux, CNRS, Inserm, Laboratoire Acides Nucléiques: Régulations Naturelle et Artificielle (ARNA, U1212, UMR5320), IECB, 2 rue Robert Escarpit, 33600 Pessac, France.

[2] Université de Bordeaux, CNRS, Inserm, Institut Européen de Chimie et Biologie (IECB, UMS3033, US001), 2 rue Robert Escarpit, 33600 Pessac, France.

[‡] Present address: CEA Saclay, DRF/JOLIOT, Service de Pharmacologie et d'Immunoanalyse, 91191 Gif sur Yvette Cedex, France

* Corresponding author. Tel. +33 (0)5 4000 2940, email v.gabelica@iecb.u-bordeaux.fr

**Abstract**

The amount of internal energy imparted to the ions prior to the ion mobility cell influences the ion structure and thus the collision cross section, and non-covalent complexes with few internal degrees of freedom and/or high charge densities are particularly sensitive to collisional activation. Here we investigated the effects of virtually all tuning parameters of the Agilent 6560 IM-Q-TOF on the arrival time distributions of Ubiquitin[7+], and found conditions in which the native state prevails. We will discuss the effects of solvent evaporation conditions in the source, of the entire pre-IM DC voltage gradient, of the funnel RF amplitudes, and will also report on ubiquitin[7+] conformations in different solvents, including native supercharging conditions. Collision-induced unfolding (CIU) can be conveniently provoked either behind the source capillary or in the trapping funnel. The softness of the instrumental conditions behind the mobility cell were further optimized with the DNA G-quadruplex $[(dG_4T_4G_4)_2 \cdot (NH_4^+)_3 - 8H]^{5-}$, for which ion activation results in ammonia loss. To reduce the ion internal energy and obtain the intact 3-$NH_4^+$ complex, we reduced the post-IM voltage gradient, but this resulted in a lower IM resolving power due to increased diffusion behind the drift tube. The article describes the various trade-offs between ion activation, ion transmission, and ion mobility performance for native MS of very fragile structures.

**Introduction**

Native mass spectrometry (native MS) involves measuring ions that have preserved a memory of the non-covalent interactions that were present in solution.[1-4] Intramolecular non-covalent interactions are responsible for the folding into specific secondary and tertiary structures, and intermolecular non-covalent interactions are responsible for the assembly of several molecules into complexes, also called "quaternary structures". The principles underlying native MS are common to mass spectrometry of intact biologically relevant molecules (proteins, nucleic acids, sugars) and of synthetic architectures[1] (artificial foldamers or supramolecular complexes). The main difference lies in the solution conditions in which the samples are dissolved for the analysis: biologically relevant complexes are studied in an electrospray-compatible buffer that best mimics the native cell environment in order to draw conclusions relevant to biology,[5] whereas artificial architectures can be studied in any medium of interest.

The experimentalist should always ensure minimal perturbation of the non-covalent interactions to be detected in the gas phase, so that the measurements provide information on the structures that were originally present in the injected solution. For a mass analysis (detection of the complexes), this means that the intermolecular non-covalent interactions must be preserved from the source to the mass analyzer. For an ion mobility (IM) analysis (characterization of the shapes via the friction with a gas when dragged by an electric field[6,7]), this means that the intramolecular non-covalent interactions must be preserved from the source to the mobility cell.[8]

Here we focused on the drift tube ion mobility instrument from Agilent (the 6560 IMS Q-TOF),[9] which like the Waters Vion IMS-QTof, the Tofwerk IMS-TOF or the Bruker timsTOF has an ESI-IM-Q-TOF configuration (as opposed to the Waters Synapt™ instruments which have an ESI-Q-IM-TOF configuration). In this configuration, both rupture of intramolecular interactions before the IM cell and rupture of the intermolecular interactions between the IM cell and the mass analyzer can affect the interpretation of ion mobility results. Pre-IM activation causes a gas-phase unfolding that could be falsely attributed to a solution unfolding.[10,11] Also, in IM calibration procedures, it is also extremely important that the calibrant ions are produced with the same distribution of collision cross sections in the calibrating instrument and in the calibrated instrument.[12] For this reason, native proteins were found to be poorer calibrants than denatured ones.[13] On the other hand, pre-IM activation can be exploited to study the gas-phase unfolding pathways, in so called collision-induced unfolding (CIU) experiments,[14] which are the IM analogue of collision-induced dissociation (CID) experiments in MS. Finally, post-IM fragmentation is often to be avoided as well, because if the arrival time distribution of a given *m/z* range is actually the arrival time distribution of the parent ion that traveled in the IM and later led to the product ion having this particular *m/z*, IM could be misinterpreted.

The degree to which non-covalent interactions can be altered in the gas phase depends on the internal energy of the ions, on the number of degrees of freedom on which this internal energy is redistributed, on the number and strength of the interactions involved, on the strength of Coulomb repulsion. Thus, complexes of smaller size and of higher charge density are more sensitive to ion activation. Here we present how we optimized the Agilent 6560 IMS Q-TOF. We used bovine ubiqutin$^{7+}$ to minimize pre-IMS unfolding and the DNA G-quadruplex $[(dG_4T_4G_4)_2 \bullet (NH_4^+)_3 -8H]^{5-}$, which undergoes ammonia loss at low internal energy,[15] to further minimize post-IMS fragmentation. Ubiquitin has frequently been used to assess IM instrument softness, because the IM profile of charge states 6+ to 8+ is extremely sensitive to internal energy effects, whether in the source,[16] in the transfer,[17] or in the pre-IM trapping region.[18,19] The solution composition also plays a role on the detected gas-phase conformations.[16,20] The native state of ubiquitin[7+] has a collision cross section of ~1000 Å² in helium (~1300 Å² in nitrogen). In contrast with a recent report on the same instrument,[21] we could obtain predominantly the native state on ubiquitin[7+], both in nitrogen and in helium drift tube conditions. For the G-quadruplex, we found that fragmentation mainly occurred after the IM when using the default instrumental settings, and when lowering the voltage gradient one had to trade-off softness for reduced IM resolution. We will describe here the rationale for optimizing the instrument for native MS of analytes bearing extremely energy-sensitive non-covalent interactions.

**Experimental**

Bovine ubiquitin, >98% purity (SDS-PAGE) was purchased from Sigma (St Quentin Fallavier, France). The oligodeoxynucleotide strand dGGGGTTTTGGGG was purchased from Eurogentec (Seraing, Belgium). The G-quadruplex was prepared by mixing 100 µM single strands in 150 mM aqueous ammonium acetate, and waiting at least 24 h. Ammonium acetate 5 M solution (molecular biology) was from Fluka. Water was nuclease-free from Ambion™ (ThermoFischer Scientific, Illkirch, France). Acetic acid glacial, formic acid 99%, and methanol absolute were from Biosolve (Dieuze, France). Sulfolane 99% was from Aldrich (St Quentin Fallavier, France).

All experiments were run on an Agilent 6560 IMS-Q-TOF,[9] equipped with the alternate gas kit, which regulates the drift tube pressure with a flow controller based on continuous readings with a capacitance gauge. The pumping system was modified in-house by using the default scroll pump for the rear of the instrument only, and connecting an ECODRY 40 Plus pump (Leybold France SAS, Les Ulis, France) to the source region. To ensure appropriate pressure differentials for collision cross section determination, the pressures were set to 3.76 Torr in the drift tube and 3.47 Torr in the trapping funnel before the flow controller was turned on. This resulted, with the flow controller on, in pressures 3.89 Torr in the drift tube and 3.63 Torr in the trapping funnel. The differential (0.26 Torr) ensured that only the intended gas (helium or nitrogen) is present in the drift tube, despite using atmospheric pressure ionization. The acquisition software version was B.07.00 build 7.00.7008. The IM data processing software IM-MS browser B.07.02 build 7.02.210.0 was used to calculate collision cross section values of well-defined peaks, and to export arrival time distributions. All graphs were constructed using SigmaPlot 12.5. For the CIU plots, because exported arrival time distributions lack data points in regions where there is no intensity; each arrival time distribution was smoothed using a negative exponential, sampling 0.02 and polynomial order 1, to produce evenly spaced data (200 intervals).

For ubiquitin in the positive mode, optimization of the pre-IMS zone was done with a 0.75 µM solution in 99% $H_2O$ and 1% acetic acid, like in the Bleiholder group did for the soft tuning of trapped IMS,[22] on the 7+ ion. Table 1 lists the parameter setpoints optimized by the manufacturer (on the Agilent tunemix ions) at the end of the installation of our instrument. We also list the parameters described by May et al. on their ubiquitin study, and the full list of parameters further optimized in the present study. The effects of changing each parameter will be described in the remainder of the paper.

*Table 1.* Tuning parameters (positive mode) in the pre-IMS zone. All voltages are floating on the ion mobility tube entrance. The parameters are transposable to the negative mode by reversing the signs.

| Parameter | Installation by Agilent | May et al.[21] | Optimized on native Ubi[7+], drift tube in He | Optimized on native Ubi[7+], drift tube in $N_2$ |
|---|---|---|---|---|
| *Source: gas temperature* | 325 °C | 25 °C | 220 °C | 220 °C |
| *Source: drying gas* | 5.0 L/min | 13 L/min | 1.5 L/min | 1.5 L/min |
| *Source: nebulizer pressure* | 20 psig | None | 9 psig | 9 psig |
| *Source: capillary* | 4000 V | Nanospray | 3500 V | 3500 V |
| *Optics 1: Fragmentor* | 400 V | | 300 V | 300 V |
| *IM front funnel: high pressure funnel delta* | 150 V | | 110 V | 110 V |
| *IM front funnel: high pressure funnel RF* | 150 $V_{p-p}$ | 80 $V_{p-p}$ | 100—180 $V_{p-p}$ | 100—180 $V_{p-p}$ |
| *IM front funnel: trap funnel delta* | 180 V | | 140 V | 160 V |
| *IM front funnel: trap funnel RF* | 150 $V_{p-p}$ | 80 $V_{p-p}$ | 160 $V_{p-p}$ | 100 $V_{p-p}$ for softness, 160 $V_{p-p}$ for CIU |
| *IM front funnel: trap funnel exit* | 10 V | | 10 V | 10 V |
| *IM trap: trap entrance grid low* | 95 V | | 70 V | 90 V |
| *IM trap : trap entrance grid delta* | 10 V | | 2 V | 2 V |
| *IM trap: trap entrance* | 91 V | | 69 V | 89 V |
| *IM trap: trap exit* | 90 V | | 67 V | 87 V |
| *IM trap: trap exit grid 1 low* | 88 V | | 64 V | 84 V |
| *IM trap: trap exit grid 1 delta* | 6 V | | 5 V | 5 V |
| *IM trap: trap exit grid 2 low* | 87 V | | 63 V | 83 V |
| *IM trap: trap exit grid 2 delta* | 10 V | | 9 V | 9 V |
| *Acquisition: Trap fill time* | 20 000 µs | | 1000 µs | 1000 µs |
| *Acquisition: Trap release time* | 150 µs | | 200 µs | 250 µs |

To optimize the post-IMS zone for softness, we used the fragile bimolecular G-quadruplex [(dGGGGTTTTGGGG)$_2$•(NH$_4^+$)$_3$-8H]$^{5-}$ ion, which readily loses ammonia upon collisional activation.[15] It is the most collision-sensitive specific nucleic acid complex we know of, and frequently use it for instrument tuning.[15] The optimized pre-IMS parameters were first transposed to the negative mode from the ubiquitin optimized ones, by changing all signs. Then, the voltages gradients behind the IMS were lowered. MS/MS mode was off (collision energy = 0 V). Table 2 lists three sets of the various parameters that can be tuned: the default parameters at installation (which give the best transmission and mobility resolution), the parameters optimized for softness for the very fragile G-quadruplex, while still preserving most of the ion signal and a decent mobility separation (although with lower resolving power, see results and discussion), and a compromise that works for most moderately fragile complexes. Further examples are given in the results and discussion.

*Table 2:* Tuning parameters (negative mode) in the post-IMS region. The "optimized" and "compromise" parameters are transposable to the positive mode by reversing the signs.

| Parameter | Installation by Agilent, suitable for intramolecular folding studies | Compromise for moderately fragile noncovalent complexes | Optimized on very fragile G-quadruplex |
|---|---|---|---|
| IM drift tube: Drift tube exit | -250 V | -210 V | -210 V |
| IM rear funnel: Rear funnel entrance | -240 V | -200 V | -200 V |
| IM rear funnel: Rear funnel RF | 150 V$_{p-p}$ | 180 V$_{p-p}$ | 180 V$_{p-p}$ |
| IM rear funnel: Rear Funnel Exit | -43 V | -35 V | -26 V |
| IM rear funnel: IM Hex Entrance | -41 V | -32 V | -24 V |
| IM rear funnel: IM Hex Delta | -9 V | -3 V | -2 V |
| Optics 1: Oct Entrance Lens | -32 V | -27 V | -21 V |
| Optics 1: Oct 1 DC | -31.3 V | -25 V | -20 V |
| Optics 1: Lens 1 | -29.4 V | -23 V | -19 V |
| Optics 1: Lens 2 | disabled | disabled | disabled |
| Quad: Quad DC | -27.8 V | -21 V | -18 V |
| Quad: PostFilter DC | -27.8 V | -21 V | -17 V |
| Cell: gas flow | 22 psi | 20 psi | 20 psi |
| Cell: Cell Entrance | -26.8 V | -20 V | -16 V |
| Cell: Hex DC | -25.8 V | -20 V | -16 V |
| Cell: Hex Delta | 9 V | 3 V | 3 V |
| Cell: Hex2 DC | -16.6 V | -14.6 V | -12 V |
| Cell: Hex2 DV | 3 V | 1.5 V | 1 V |
| Optics 2: Hex3 DC | -13.2 V | -12.9 V | -11 V |
| Extractor: Ion Focus | -10 V | -10 V | -10 V |

**Results and discussion**

Tuning is a multiparametric and iterative process. To improve softness, the general reasoning is to reduce the ion acceleration by electric fields to reduce the laboratory frame collision energy between the ion and the background gas. However, there is often a compromise between softness and ion transmission. Also, too much softness can mean insufficient desolvation or declustering. Finally, the actual effect of a given voltage difference depends on the gas pressure and flows.[23,24] We will show the effect of each type of parameter either side of our final optimized values to illustrate what happens when departing from this optimum. It must be understood however that there may exist different sets of parameter which could constitute another optimum. We don't claim we have reached the global optimum, but we have reached one that suits our applications on native MS of ~10 kDa complexes and very fragile complexes, including for performing collision-induced unfolding experiments.

*Tuning the instrument pre-IM region for softness*

In the pre-IM region, we examined twenty different parameters. Figure 1 depicts the optimized DC voltage gradient from the end of the glass capillary to the IM entrance. Supporting information Figure S1 compares visually the DC gradient in the default instrument tuning (which we presume is similar to that used by May et al.[21] unless otherwise specified in Table 1) and in the optimized tuning for the drift tube operated in helium. As detailed below, the main difference lies in the voltage gradients in the ion trapping region.

The voltage called "fragmentor", applied at the end of the glass capillary, was already known to influence softness. However, the tuning of the other parameters was less intuitive, and we found one voltage difference (between the trap entrance and the trap entrance grid) that unexpectedly had a large influence on the results. All parameters are explained in detail below. We focus here on describing the tuning for softness with the drift tube in helium, which was more sensitive than in nitrogen, and will mention any differences found in nitrogen.

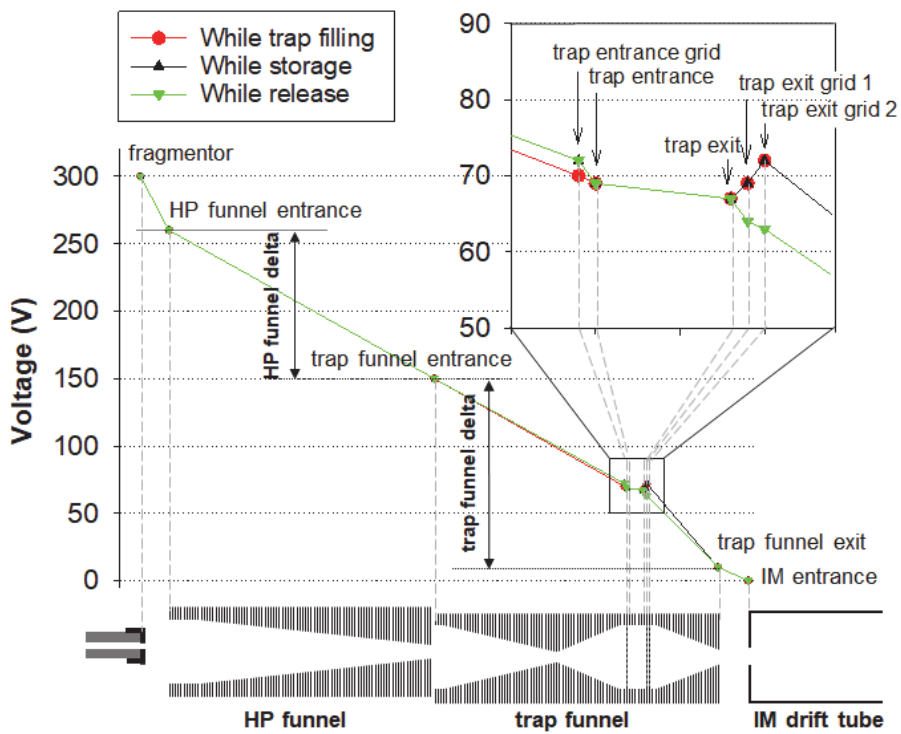

*Figure 1*: DC voltage profile before the ion mobility (IM) drift tube entrance, with parameters optimized for 3.89 Torr helium in the drift tube and a pressure differential of 0.26 Torr compared to the trapping funnel. The parameters are listed in Table 1.

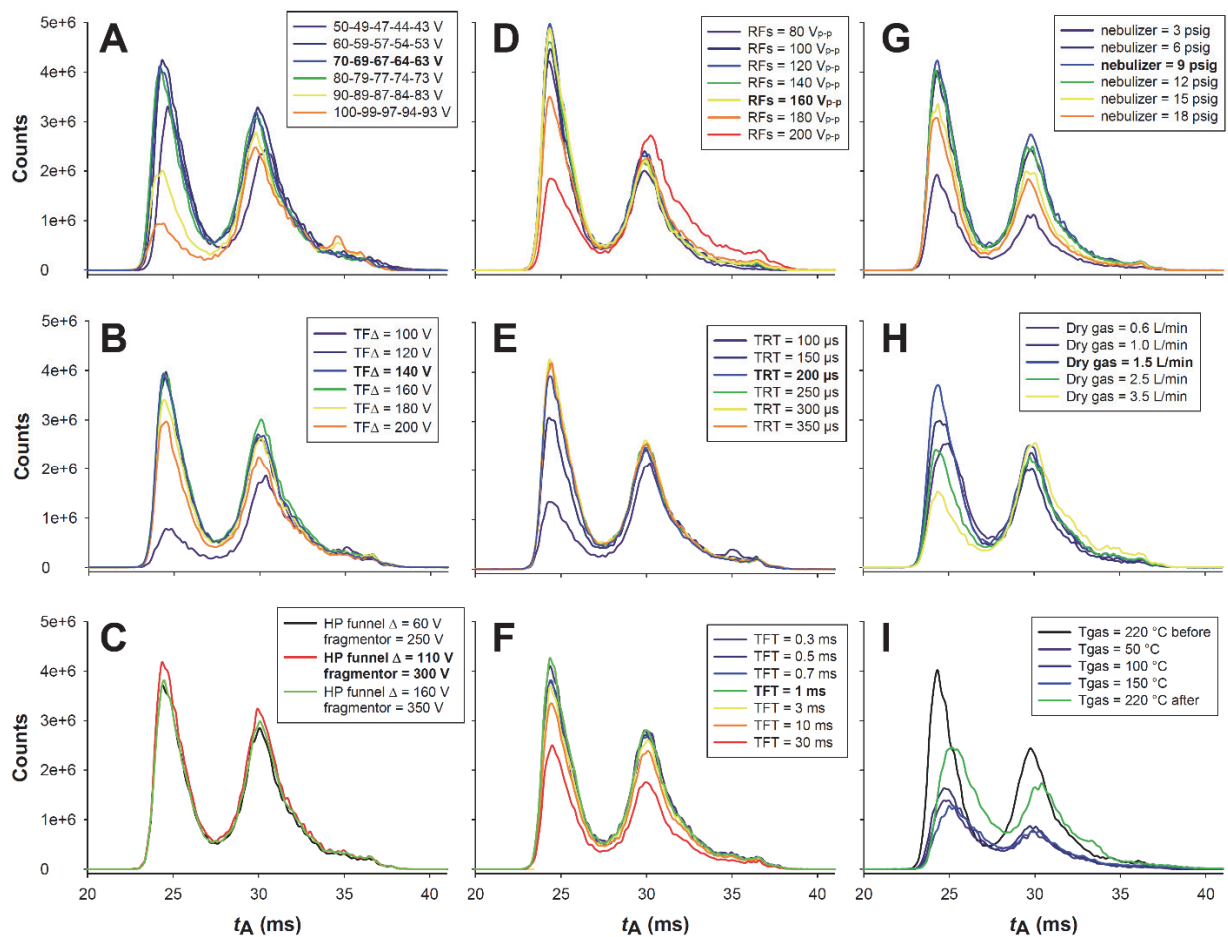

*Figure 2*: Influence of various tuning parameters on the arrival time distribution of Ubiquitin[7+], sprayed from a 0.75 µM solution in 99% $H_2O$ and 1% acetic acid, with the drift tube operated in 3.89 Torr helium. The same number of scans were summed in all cases. The drift tube was operated with $\Delta V$ = 390 V. The peak at ~25 ms corresponds to the native form ($CCS_{He}$ = 981 Å²; standard deviation was 7 Å² from 10 independent measurements). The optimum value is in bold. Effects of A) voltages on the whole trapping region (see inset of Figure 1), the values written being the trap entrance grid low-trap entrance-trap exit-trap exit grid 1 low-trap exit grid 2 low; B) trap funnel delta; C) HP funnel RF and front funnel RF, both set at the same value; D) trap release time; E) trap fill time; F) HP funnel delta and fragmentor voltage; G) electrospray nebulizer pressure; H) dry gas flow; I) dry gas temperature effect, in order of recording. Note that after lowering the temperatures for a while, the system takes a very long time to recover its initial condition.

***DC level of the trapping region***. All DC voltages are floating on the IM entrance, which is thus defined here as zero volts. The trapping funnel is located before the IM entrance. The trapping funnel has an hourglass shape, with an entrance, an exit, and the trapping region delimited by grids. Further towards the source we find the high pressure funnel (which has entrance and exit voltages), and the "fragmentor", which accelerates ions from the glass capillary towards the HP funnel. As shown in Figure 1 and Table 1, on the data acquisition software for some voltages one enters absolute values, and for others one enters voltages differences ("delta"s). The trap funnel exit is optimum at 10 V in all cases. The voltage levels of the trapping region of the trapping funnel (in the 70-volts range on Figure 1) will influence the acceleration energy of ions released from the trap and injected in the drift tube.

In Figure 2A, we show that this "injection energy" influences the softness. In helium, when the trap DC voltages are in the 90-volts or 100-volts range, such as in typical installation parameters for nitrogen, the conditions are less soft and the peak of the native conformation at ~25 ms disappears to the profit of more extended (longer drift time) conformations. Too low trap DC voltages are also detrimental, because the IM peaks are shifted to longer times. The reason is that the ions travel slower to the IM: this delay contributes to the dead time $t_0$ (time spent outside the drift region), and the conditions are detrimental to the resolving power. In helium, the optimum trap DC level is thus in the 70-vols range. In nitrogen, the activation effects are less acute (see supporting Figure S2A), and the 90V-range is optimum, because in the 70V-range the ions drift too slowly towards the IM. The influence of other voltages in the trapping region will be discussed in the section on CIU. The parameters listed in Table 1 are the softest ones that do not compromise ion trapping efficiency or IM resolution.

***Funnel DC gradients***. Continuing towards the source, we have the trap funnel entrance, which voltage is defined by the trap funnel exit voltage plus the trap funnel delta. Figure 2B shows the influence of the trap funnel delta. At high values (200 V is the maximum allowed), the conditions become slightly activating, and one loses signal intensity. But counterintuitively, at low voltages (100 V) the conditions are also more activating, and the signal is lower. A possible explanation is that if the ions move too slowly between the funnel entrance through the hourglass to the trapping region, they wander around longer before entering in the trapping region and have time to be activated by the RF and change conformation. Our results therefore suggest that ion trapping in the first part of the funnel can occur at too low voltages. The optimum front funnel delta voltage was set to 140 V in helium (160 V in nitrogen, in line with the trap level which is also 20 V higher, see Figure S2B). In contrast, the high-pressure funnel delta shows no such effects, and any value of HP funnel delta between 60 V and 160 V gives similar results (Figure 2C). A possible explanation is that the gas flows play a larger role than the voltage gradient on the ion transport in this region, and thus trapping does not occur. The fragmentor voltage defines an acceleration region at the entrance of the high-pressure funnel, and its effect will be described in the section on CIU. For soft conditions, its value is set slightly above the HP funnel entrance absolute value.

***Trapping funnel RF amplitude***. The high-pressure funnel and trap funnel have also radiofrequencies (1.5 MHz for the high-pressure funnel, 1.2 MHz for the trapping funnel). May et al. described that both RF amplitudes must be decreased to 80 $V_{p-p}$ to have softer conditions. We reproduced this trend when operating the drift tube in nitrogen (Figure S2C), and chose 100 $V_{p-p}$ to optimize the signal. However, in helium, the situation is different: from 80 $V_{p-p}$ to 160 $V_{p-p}$, no significant ion activation was observed. Unfolding is only noticed at 180 $V_{p-p}$ and higher (in nitrogen, unfolding is noticeable already at 140 $V_{p-p}$). The reason for the difference is that when helium is used in the IM tube, helium is also injected in the funnels to avoid nitrogen contamination in the tube, and ions move faster overall. Thus, in helium, similar RF amplitudes have a lower effect on ion activation than in nitrogen.

It is surprising that the trap funnel RF amplitude has a greater effect on ion activation in nitrogen than in helium, while the DC gradients in the same region have a greater effect in helium than in nitrogen. A possible origin of the difference could be a different positioning of the ion cloud in the trapping region, depending on the ion gas, for given RF amplitudes. If radial ion placement differs according to the gas composition, it is possible that a given DC voltage different has different effects for the ions in terms of actual electric field. However, to test whether this hypothesis explains the observations, simulations of ion trajectories in presence of collisions with the background gases would be required.

We will describe below that it is possible to carry out CIU experiments in the trapping region. To do that, however, the trap funnel RF amplitude has to be high enough, presumably to confine the ions in the region where the DC gradient is applied. Also, for larger ions the RF amplitudes have to be increased (for example, 200 $V_{p-p}$ is optimum for native bovine serum albumin (66 kDa) transmission, data not shown). The final choice of the RF amplitudes will depend on the type of experiment: 100 $V_{p-p}$ for native MS on fragile molecules in nitrogen, and 160 $V_{p-p}$ or higher in helium and for CIU. We also found that the HP funnel RF amplitude had little effect on transmission and softness (Figure S3): 100 $V_{p-p}$ and 180 $V_{p-p}$ give the same results. Again, gas flows may dictate ion transport more than voltages in the high-pressure funnel.

***Trapping and release times***. We noticed an effect of the trap release time (the time during which the trap voltages are set to the green ramp in Figure 1, to push the ions towards the IM) on the arrival time distribution. Short trap release times are supposed to help bunch the ion packet towards the IM with a short time spread (contributing minimally to the width of $t_0$). However, with trap release times shorter than 200 µs, ubiquitin gets more activated and the total ion signal decreases (Figure 2E). A possible explanation is that the fraction of the ion population that had the opportunity to exit the trap during short release times had to do so quickly, and thus traveled faster and got activated more. Longer times than 200 µs did not change the allure of the peaks. The same effect was observed in nitrogen, and the optimum for softness was 250 µs. Again, the difference comes from the slower movement of ions in nitrogen. The trap fill time (Figure 2F) did not influence the softness. However, the ion signal decreases at long trap fill times. Normally the ion signal should increase linearly with the fill time unless there is space charge, as described on a similar trap with one entrance grid and two exit grids.[25] The same phenomenon was observed on a solution ten times less concentrated (0.075 µM, see Figure S4), giving ten times lower signal with the standard ESI source, so the phenomenon is not due to space charge. The origin of this phenomenon is unknown.

### Influence of the electrospray source parameters

We also examined how the electrospray source parameters influence the activation of ubiquitin[7+]. Here the effects of the electrospray process (at stake in the intermediate phase consisting of the dense solvent plume) and of ion activation (desolvated ions undergoing collisions) may be combined. The nebulizer gas flow had no effect on the apparent activation. It had only an effect on the ion signal and the optimum was at 9 psig. This parameter should be optimized based on ion signal for each solvent.

In contrast, the drying gas flow had an influence on the apparent ion activation. In the standard electrospray source, a heated drying gas is coming from the entrance capillary, countercurrent to the spray. At 220 °C, high flows cause substantial activation, and low flows resulted in lower ion signals. The optimum drying gas flow, for ubiquitin infused at 190 µL/hour, was 1.5 L/min.

The drying gas temperature, when decreased, also resulted in decreased ion signals, without necessarily decreasing the apparent ion activation. The droplets interact only very transiently with the heated nebulizer gas and undergo endothermic evaporation. Consequently, the effect of gas heating is not equivalent to heating the solution, and gas temperatures higher than the solution melting temperature can be used. However, too low temperatures were very detrimental to the IM signals: over a time scale of several minutes, the IM peaks started to shift to longer drift times, and even when going back to the usual temperature of 220 °C, the shift persisted for over an hour. Our interpretation is that when the drying gas temperature is too low, significant amounts of electrospray solvent enters the funnels and eventually contaminate the drift tube, where even trace amounts of polar molecules in the helium bath gas slow down the ions. Given that the drying gas temperature did not affect the apparent harshness of the conditions, we recommend using sufficiently high temperatures in electrospray, to avoid solvent contamination in the pressure differential. We found that solvent-dependent contamination problems in the tube are more likely to occur when the drift tube is operated in helium, compared to nitrogen.

Finally, we compared three different ion sources: the standard ESI source (results above), the JetStream™ source (Agilent Technologies), and static nanospray ionization using extra coated L borosilicate capillaries (ThermoFischer Scientific, Illkirch, France). Figure S5 shows the comparison for Ubiquitin$^{7+}$, with nitrogen as drift gas. The standard ESI source appears slightly softer. Similar conclusions were reached by Konermann's group on a Waters Synapt ion mobility mass spectrometer.[13]

*Comparison of ubiquitin sprayed from different solvents*

To compare ubiquitin sprayed from different solvents, we thus used the drift tube in nitrogen. After starting the injection of each solvent mixture for at least five minutes, we controlled the pressure differential with the flow controller off then on again, and then recorded the data presented here. The optimum source and transfer conditions of Table 1 were used for all solvent mixtures. The resulting arrival time distributions are shown in Figure 3.

Figure 3A shows the arrival time distribution obtained in water/methanol and 0.2% formic acid, i.e. the denaturing conditions used by May et al on the Agilent 6560.[21] Even with the optimized soft instrument tuning, the conformations under this charge state are mostly denatured. In contrast, when using 1% acetic acid instead of 0.2% formic acid (Figure 3B), which are the conditions used by Russell's group on the Waters Synapt,[26] despite the charge state distribution is also centered on high charge states, the conformations under the 7+ charge states are predominantly compact, which may be due to a small fraction still with a native fold in solution. The apparent discrepancy between the results of May et al. and Russell et al. was thus mainly due to the solvent choice.

In the interest of native MS, we also investigated native supercharging conditions, here 0.5% sulfolane in 150 mM $NH_4OAc$ (Figure 3C). The 7+ ion is abundant and its arrival time distribution is mainly constituted from a compact conformation, and the conditions appear even softer than in $H_2O$/acetic acid (Figure 3F). However, the peak is slightly offset to a higher collision cross section (1337 Å²) compared to all sample preparations without sulfolane (1262 Å², standard deviation: 9 Å²). In comparison, the compact peak obtained from 50 mM $NH_4OAc$ (i.e., Russell's native conditions,[26] Figure 3D) or 20 mM $NH_4OAc$ (Figure 3E) is similar to the one obtained from aqueous acetic acid (i.e., May's native conditions,[21] Figure 3F). The supercharging mechanisms in the presence of $NH_4OAc$ are still a subject of debate,[27-30] and our results suggest that although the conformations are compact, they are not necessarily identical to the native ones. These preliminary results call for further systematic studies, which are beyond the scope of this paper and will be reported elsewhere.

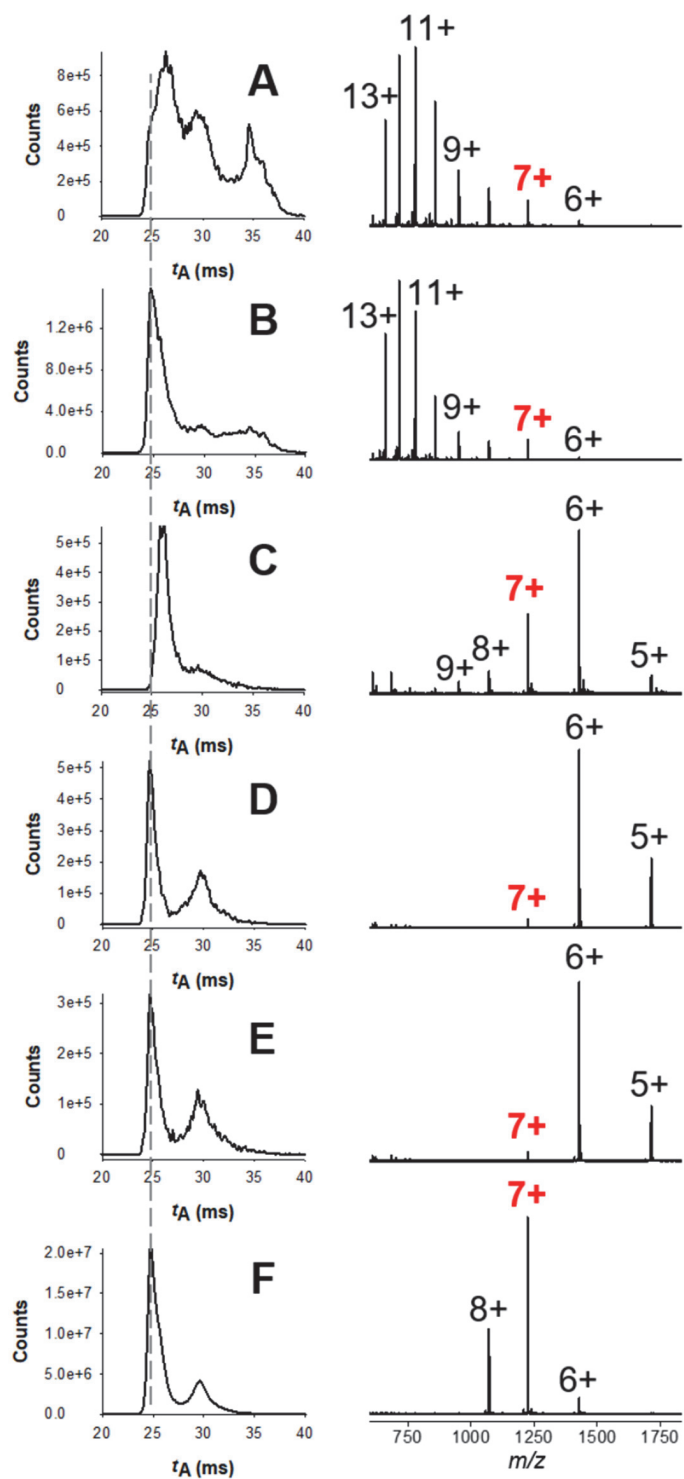

*Figure 3*: Arrival time distribution of the ubiquitin[7+] and charge state distributions obtained from different solvents (drift tube in 3.89 Torr nitrogen, $\Delta V$ = 1390 V): A) $H_2O$/MeOH/formic acid (49.9/49.9/0.2 % vol); B) $H_2O$/MeOH/acetic acid (49.5/49.5/1 % vol); C) 150 mM $NH_4OAc$, 0.5% sulfolane; D) 50 mM $NH_4OAc$; E) 20 mM $NH_4OAc$; F) $H_2O$/acetic acid (99/1 % vol).

*Collision-induced unfolding (CIU) experiments*

We found two parameters which could cause substantial activation: the fragmentor voltage and the trap entrance grid delta. The fragmentor is accelerating the ions between the extremity of the coated glass capillary and the entrance of the high pressure funnel. Figure 4A shows the resulting CIU plot: at 450 V the abundance of the intermediate state becomes slightly larger, and at 490 V an extended state appears.

Figure 4B shows that CIU plots can also be obtained by varying the trap entrance grid delta, which is the voltage difference between the trap entrance grid "low" (during trap filling), and "high" (during storage and release, see Figure 1). This means that, after filling, the ions in the trap are subjected, on the entrance side, to a sudden voltage difference equal to (trap entrance grid delta + trap entrance grid low – trap entrance), and that this voltage can activate the entire ubiquitin$^{7+}$ population to unfold it. Note that for carrying out CIU experiments using the trap entrance grid delta voltage, the trap funnel RF amplitude has to be increased to 160 $V_{p-p}$. Too low trap RF amplitudes result in signal losses at some voltages, and aberrant CIU plots if normalization is done spectrum per spectrum (see Figure S6).

Comparing Figures 4A and 4B also reveals that the activation processes are not equivalent in the two regions. When ubiquitin$^{7+}$ is activated in the trap, three distinct extended conformations are produced, eventually two at the highest voltages. This is in line with previous reports by IMS-IMS,[17,20] CIU in a Waters Synapt,[31] and by activated TIMS.[32] When activated in the fragmentor region, however, a single extended population is observed, which is broader and at a median arrival time compared to the conformers found by activation in the trap. We conjecture that residual solvent vapor may still be present in the high pressure funnel where the fragmentor operates, and that ion-molecule reactions could contribute rearranging the extended conformers. This hypothesis would warrant further investigation, depending on source conditions and on the solvent, and if confirmed, comparing CIU in the fragmentor and in the trap could be a useful method to compare CIU in partially solvated vs. gas phase environments.

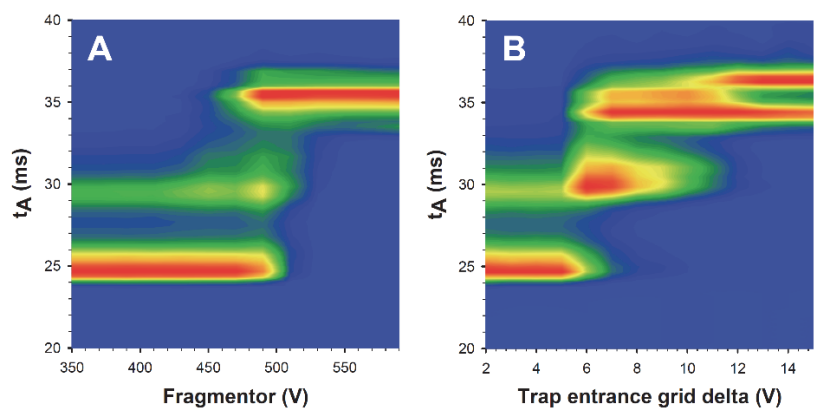

*Figure 4*: Collision-induced unfolding (CIU) of ubiqutin$^{7+}$, with the drift tube in 3.89 Torr nitrogen (optimized voltages from Table 1, $\Delta V$ = 1390 V), obtained by changing (A) the fragmentor voltage (trap funnel RF amplitude = 100 $V_{p-p}$), or (B) the trap entrance grid delta voltage (trap funnel RF amplitude = 160 $V_{p-p}$).

*Optimization of the post-IM region for fragile non-covalent complexes*

When fragmentation of very fragile complexes must be reduced also behind the drift tube, the voltage gradients from the rear funnel down to the TOF region can be lowered. After choosing the softest parameters before the IM, we optimized the post-IM region using the ammonium-bound bimolecular G-quadruplex [(dGGGGTTTTGGGG)$_2$•(NH$_4^+$)$_3$-8H]$^{5-}$, which served to benchmark several instruments,[15] and the results are shown in Figure 5. In addition to the default parameters provided by the manufacturer following installation, we devised two sets of voltage gradients: one is the softest that is still compatible with IMS for our complexes, while minimizing fragmentation ("very fragile" tuning parameters), and one that is a compromise giving higher transmission and better IM resolution. The voltage gradients are shown in Figure 5A. Again, some of the applied voltages are differences between an entrance and a delta voltage entered in the data acquisition software. By lowering the entire voltage gradient, we could obtain the intact three-ammonium complex as the major species when the drift tube is in helium (Fig. 5D), and even the five-ammonium complex, which is also specific,[15] with the drift tube in nitrogen (Fig. 5G).

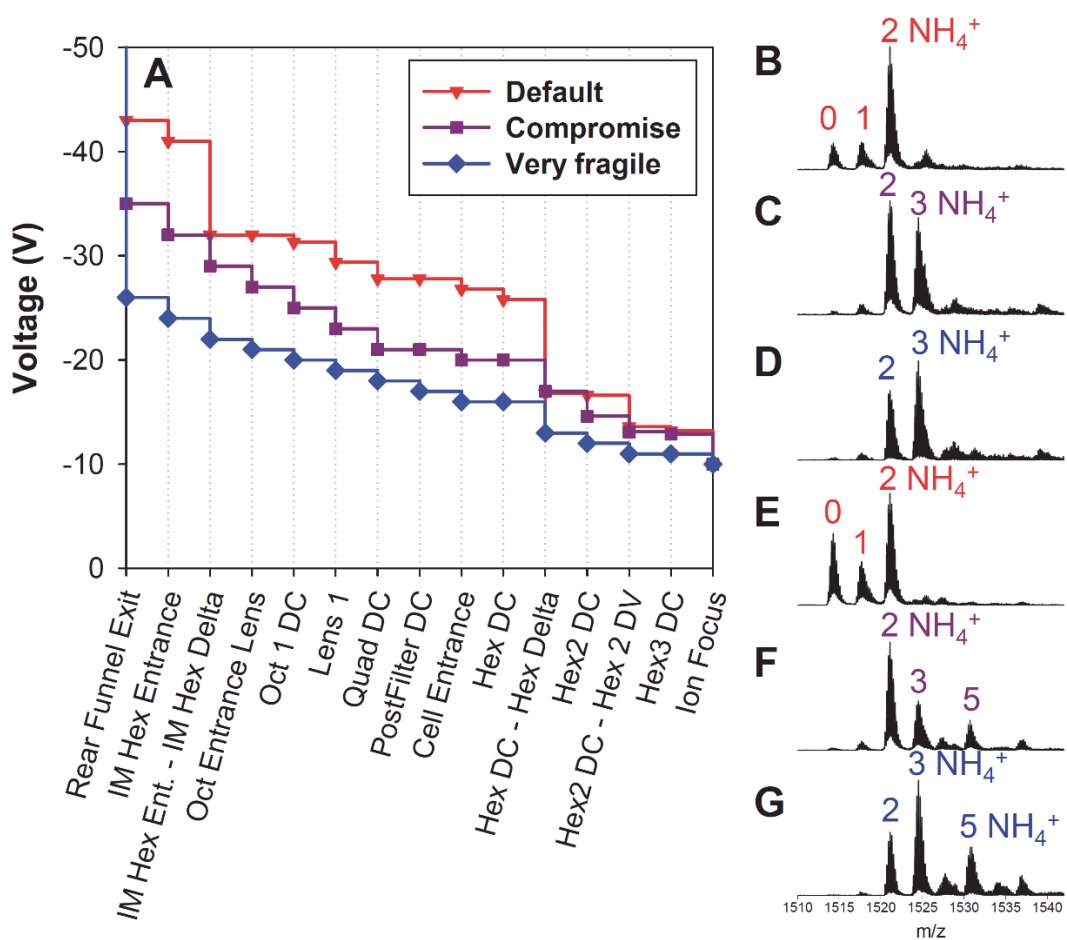

*Figure 5*: (A) Illustration of the DC voltage gradient behind the drift tube in negative mode, with the three sets of voltages listed in Table 2. (B—G) ammonium adduct distribution on the bimolecular G-quadruplex [(dG$_4$T$_4$G$_4$)$_2$•(NH$_4^+$)$_n$–(5+n)H$^+$]$^{5-}$, obtained in helium (B—D) with the three different gradients (B: default, C: compromise, D: very fragile) and in nitrogen (E—G, same color code).

The choice of the voltage set depends on the application. If post-IM fragmentation is to be avoided at all costs, the shallowest possible gradient is recommended, but this comes at the expense of ion signal and IM resolution, as shown in Figure S7 on ubiquitin[7+]. With too shallow gradients, the ions wander around longer, so the time spent outside the IM drift tube ($t_0$) increases, and diffusion increases. With extremely shallow gradients, one may even reach a limit where the ions are trapped in hexapoles, and all IM resolution is lost (MS analysis is still feasible). For studying purely intramolecular conformations (provided that no multimer dissociation is interfering), the default steeper gradients should be used, to maintain the best possible IM resolution.

**Conclusions**

Optimizing the Agilent IMS-Q-TOF for native MS, especially when the drift tube is operated in helium, required departing significantly from the recommended default settings. After examining systematically the combined effects of nearly forty parameters, we obtained conditions allowing to obtain mainly the native ubiquitin[7+] conformation, and the intact fragile G-quadruplex [(dGGGGTTTTGGGG)$_2$•(NH$_4^+$)$_3$-8H]$^{5-}$. The main difference between our IM profiles and those reported by May et al.[21] come from (1) the solvent used and (2) the entrance grid potential upon ion release. The voltage difference between the trap entrance grid and the trap entrance is a key parameter to tune ion activation in the trapping region.

Some tuning parameters differ in helium and in nitrogen because the ions move faster in helium, and slightly softer conditions could be obtained in nitrogen. We also demonstrated that collision-induced unfolding (CIU) experiments could be conducted in two different ways: by increasing the fragmentor voltage or the trap entrance grid delta voltage. Both types of CIU data are interesting, given that the fragmentor operates in a region where solvent vapors may still be present, whereas the trapping region is closer to pure gas-phase conditions. Understanding all these parameters will be useful for several kinds of native MS applications, for example to choose conditions that minimally perturb the structures coming from the solution, to activate them intentionally, or to minimize fragmentation both before and after the IM in order to correctly assign the mobility peaks. In future work, the optimized tuning described herein can be exploited to investigate particularly fragile complexes, for example maintained by only a few weak non-covalent interactions, or with high charge densities and thus higher Coulomb repulsion between subunits.

**Acknowledgements**

The research leading to these results has received funding from the European Research Council under the European Union's Seventh Framework Programme (FP7/2007-2013) / ERC grant agreement n° 616551 (project DNAFOLDIMS to VG).

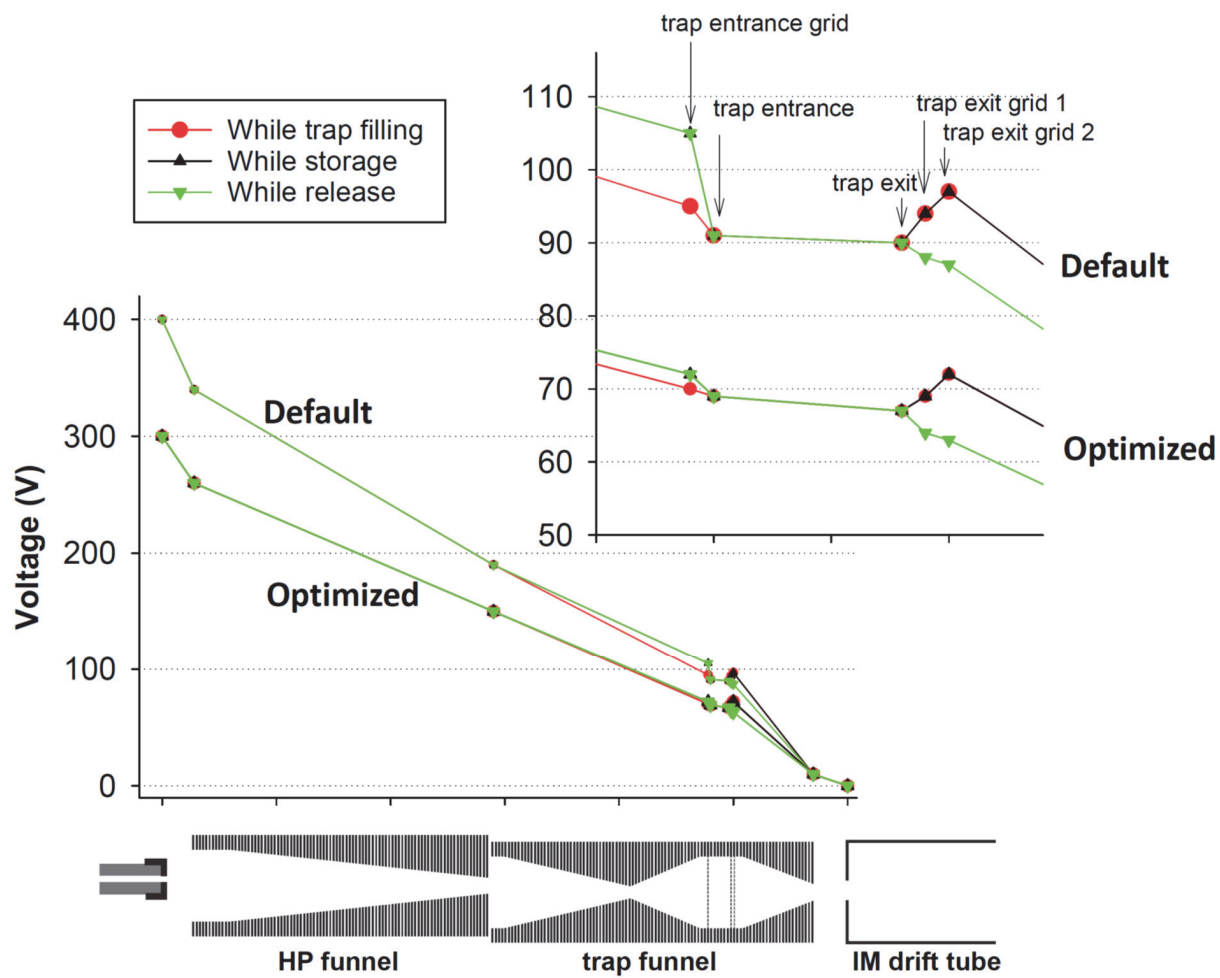

*Figure S1*: Comparison of the pre-IM DC voltage gradient in the default tuning parameters and in the tuning parameters optimized for soft conditions when the drift tube is operated in 3.89 Torr helium and the trapping funnel pressure is 3.73 Torr.

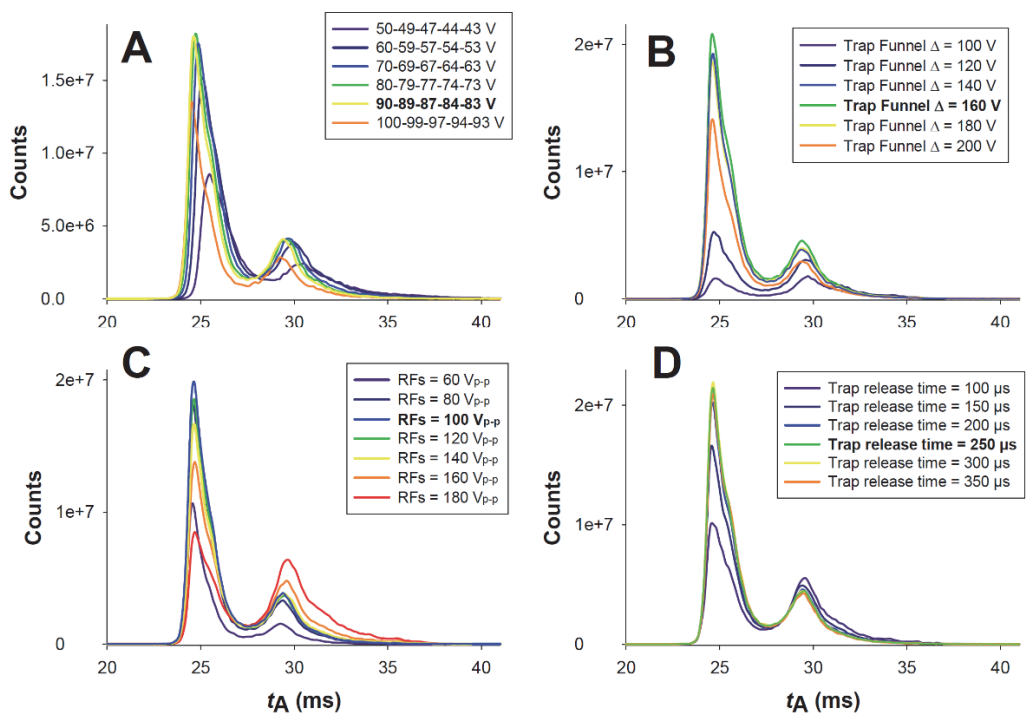

*Figure S2:* Influence of tuning parameters on the arrival time distribution of Ubiquitin$^{7+}$, sprayed from a 0.75 µM solution in 99% H2O and 1% acetic acid, with the drift tube operated in 3.89 Torr **nitrogen**. The drift tube was operated with ΔV = 1390 V. The peak at ~25 ms corresponds to the native form (CCS$_{N2}$ = 1262 Å²; standard deviation was 9 Å² from 5 independent measurements in the optimum conditions, which are indicated in bold). Effects of A) voltages on the whole trapping region (see inset of Figure 1), the values written being the trap entrance grid low-trap entrance-trap exit-trap exit grid 1 low-trap exit grid 2 low; B) trap funnel delta; C) HP funnel RF and front funnel RF, both set at the same value; D) trap release time.

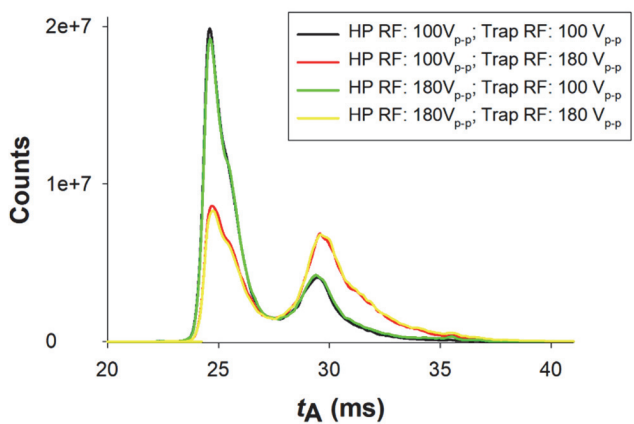

*Figure S3*: Influence of the high pressure funnel RF amplitude and trap funnel RF amplitude on the arrival time distribution of ubiquitin$^{7+}$ ions, sprayed from a 0.75 µM solution in 99% H2O and 1% acetic acid, with the drift tube in nitrogen. Only the trapping funnel RF amplitude is responsible for the ion activation.

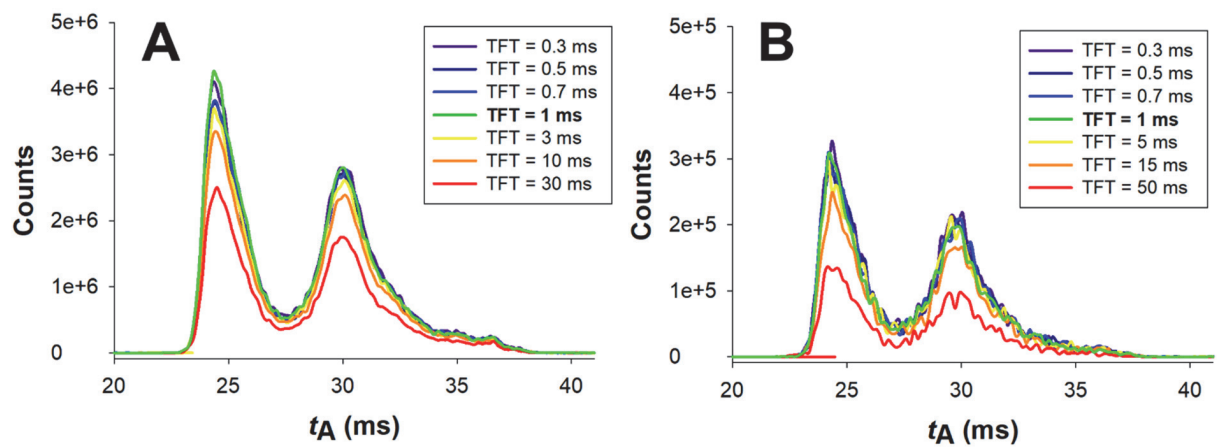

*Figure S4:* Influence of the trap fill time on the arrival time distribution of ubiquitin[7+] ions, sprayed from (A) a 0.75 µM or (B) a 0.075 µM solution in 99% H2O and 1% acetic acid, with the drift tube in helium. The same number of scans were summed in each case.

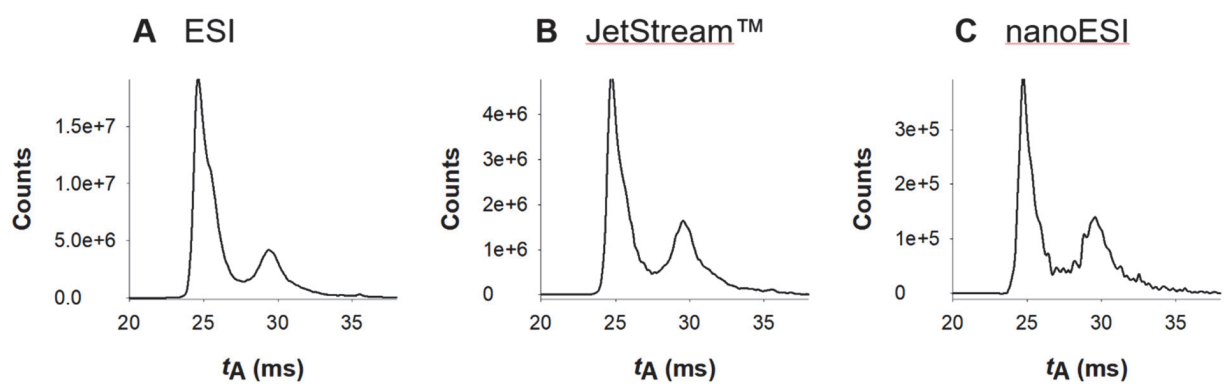

*Figure S5*: Arrival time distribution of ubiquitin$^{7+}$, sprayed from a 0.75 µM solution in 99% H2O and 1% acetic acid, with optimization instrumental parameters (see Table 1) and the drift tube in 3.89 Torr nitrogen. (A) Standard ESI source, with drift gas at 1.5 L/min and 220 °C, nebulizer gas at 9 psi, spray voltage at 3500 V. (B) JetStream™ source, with with drift gas at 1.5 L/min and 220 °C, nebulizer gas at 9 psi, spray voltage at 3500 V, jet gas flow at 2 L/min and 30°C, and nozzle voltage at 1000 V. (C) Nanospray source with manually cut coated borosilicate capillary, spray voltage at 1100 V, no nebulizer gas, and drying gas at 13 L/min and 30 °C (similar to May et al.[1]). Note that the nanospray cone diverts most of the drift gas flow.

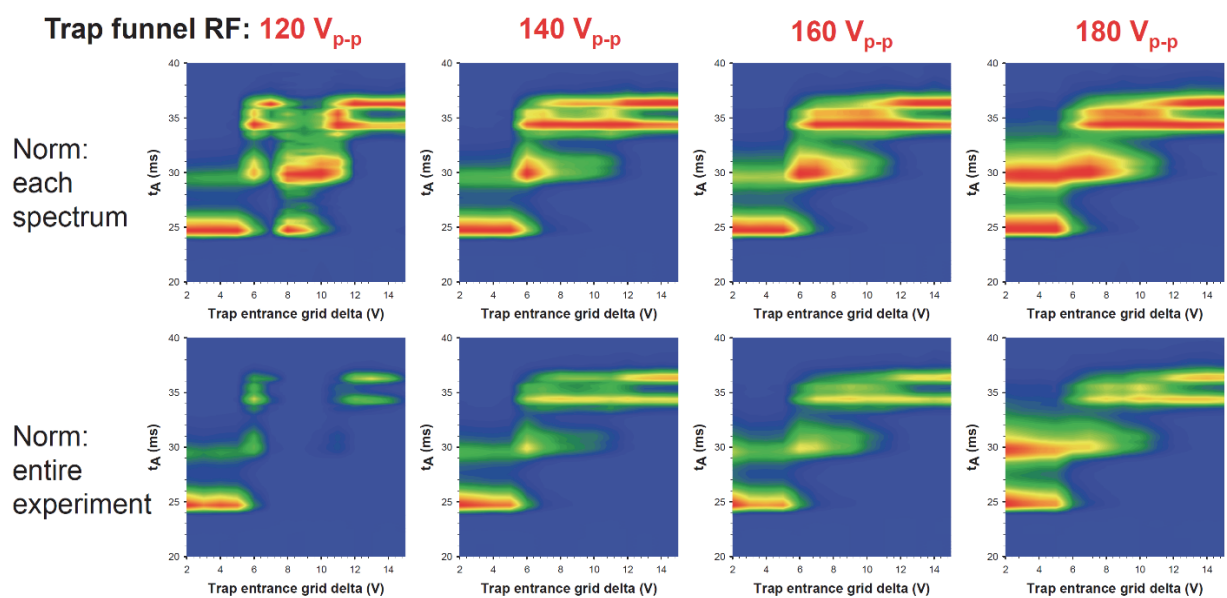

*Figure S6*: Influence of the trap funnel RF amplitude on the CIU results obtained when varying the trap entrance grid delta voltage. When normalizing the arrival time distribution at each voltage, the CIU plots look strange at low trap funnel RF (120 V$_{p-p}$). In reality, the signal intensity is lost at some specific voltage ranges (see bottom row where the normalization is done over the entire experiment, not spectrum per spectrum). As the RF amplitude is increased, there is no more signal loss, but the activation at low voltage is becoming prominent. For ubiquitin7+, 160 Vp-p presents a good comprmise to carry out CIU experiments.

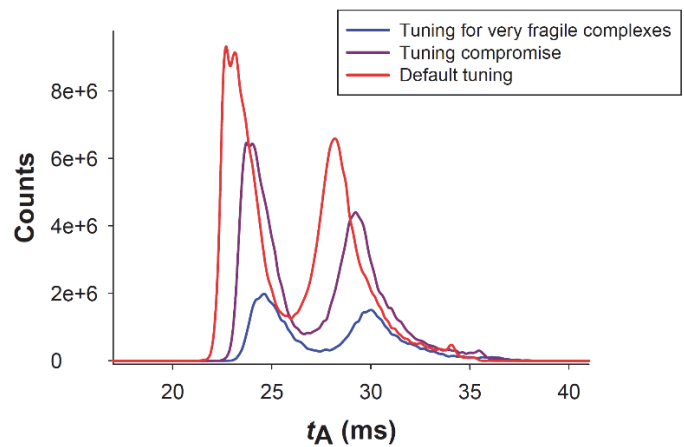

*Figure S7*: Effect of changing the voltage gradient behind the IM drift tube (parameters, see Table 2) on the arrival time distribution of ubiquitin$^{7+}$ (produced from 99% $H_2O$, 1% acetic acid, drift tube in 3.89 Torr helium with $\Delta V$ = 390 V). The measured collision cross section is unchanged, but the $t_0$ value changes (5.21 ms for the softest conditions, 4.83 ms for the compromise tuning, 3.70 ms for the default installation tuning).